\documentclass[aps,icp,twocolumn,groupedaddress]{revtex4-1}
\usepackage{graphicx,amsmath,amssymb,color}
\begin{document}

\title{Does the Sastry transition control cavitation in simple liquids?}
\author{Caitlin M. Gish}
\author{Kai Nan}
\author{Robert S. Hoy}
\email{rshoy@usf.edu}
\affiliation{Department of Physics, University of South Florida}
\date{\today}
\begin{abstract}
We examine the Sastry (athermal cavitation) transitions for model monatomic liquids interacting via Lennard-Jones as well as shorter- and longer-ranged pair potentials.
Low-temperature thermodynamically stable liquids have $\rho < \rho_S$ except when the attractive forces are long-ranged.
For moderate- and short-ranged attractions, stable liquids with $\rho > \rho_S$ exist at higher temperatures; the pressures in these liquids are high, but the Sastry transition may strongly influence their cavitation under dynamic hydrostatic expansion.
The temperature $T^*$ at which stable $\rho > \rho_S$ liquids emerge is $\sim 0.84\epsilon/k_B$ for Lennard-Jones liquids; $T^*$  decreases (increases) rapidly with increasing (decreasing) pair-interaction range.  
In particular, for short-ranged potentials, $T^*$ is above the critical temperature.
All liquids' inherent structures are isostructural (isomorphic) for densities below (above) the Sastry density $\rho_S$.
Overall, our results suggest that the barriers to cavitation in most simple liquids under ambient conditions where significant cavitation is likely to occur are primarily vibrational-energetic and entropic rather than configurational-energetic.
The most likely exceptions to this rule are liquids with long-ranged pair interactions, such as alkali metals.
\end{abstract}
\maketitle

\section{Introduction}

Repulsive forces dominate liquids' structure over a wide range of pressures and temperatures, allowing many of their properties to be understood using ``universal'' models.\cite{dyre16}
On the other hand, attractive forces become increasingly important near freezing and vaporization transitions, particularly for dynamic phenomena.\cite{dyre16,berthier09,tong20}
In general, varying the range and shape of the interatomic pair potential $U(r)$ profoundly alters both the cluster-level structure and the macroscopic properties of liquids.\cite{doye96}
Such effects can be understood in terms of the energy landscape (EL).\cite{stillinger82}
Hard-core-like repulsions and short-range attractions produce rough ELs with many basins, while softer repulsions and longer-range attractions produce opposite trends.\cite{wales04}

One ubiquitous phenomenon for which the details of attractive interactions are particularly important is cavitation, the formation of gas bubbles within liquids experiencing tensile stress.
Cavitation is entirely absent in systems with purely repulsive or short-range-attractive interactions because these systems lack distinct liquid and gas phases. 
Classical nucleation theory (CNT) performs particularly poorly for cavitation, typically underestimating the bubble nucleation rate by many orders of magnitude.
Oxtoby and Evans argued\cite{oxtoby88} that the disagreement between CNT and nonclassical nucleation theories (as well as experiments) gets progressively worse as the the range of interparticle attractions increases because CNT's assumption that bubbles are homogeneous (i.e.\ have a uniform density and pressure) becomes increasingly inaccurate.
However, few other studies have systematically examined how cavitation phenomenology varies with the range and shape of $U(r)$.

Cavitation under typical real-world conditions is inhomogeneous; it tends to nucleate at impurities, especially gaseous impurities.\cite{appel70}
Homogeneous cavitation commonly occurs in sonicated liquids\cite{finch64,maris00} and behind shock fronts.\cite{dewey18}
On the other hand, simulations have suggested that even in macroscopically homogeneous systems, cavitation preferentially nucleates in regions with lower density\cite{sastry97} and/or higher temperature.\cite{wang09}  
These results imply that accurately predicting inhomogeneity within the bulk liquid phase is a prerequisite to developing quantitatively accurate microscopic theories of cavitation.
Developing such theories is very difficult because it requires detailed knowledge of liquids' ELs' topographies, which are chemistry-dependent.\cite{doye96}

Accordingly, many of the recent advances in our theoretical understanding of cavitation have come from particle-based simulations.
Most studies of homogeneous cavitation have employed Lennard-Jones pair interactions.\cite{sastry97,wang09,baidakov11,baidakov14,baidakov16,angelil14,kinjo98,baidakov05,kuskin10,cai16}
Some of these have focused on comparison of equilibrium simulation results to various classical and nonclassical theories, for various  thermodynamic state points (various densities $\rho$ and temperatures $T$).\cite{sastry97,wang09,baidakov11,baidakov14,baidakov16,angelil14} 
Others have examined dynamic cavitation under hydrostatic expansion from a single $(\rho,T)$.\cite{kinjo98,baidakov05,kuskin10,cai16}.

Sastry \textit{et al.}\ demonstrated the existence of a cavitation transition in liquids' \textit{energy landscapes}.\cite{sastry97}
The Sastry density $\rho_S$ is the density $\rho$ at which the pressure  $P_{\rm IS}(\rho)$ within liquids' inherent structures (IS) is minimal.
For $\rho > \rho_S$, liquids' IS are homogeneous and mechanically stable:\ $\partial P_{\rm IS}/\partial \rho > 0$.
For $\rho < \rho_S$, liquids' IS consist of coexisting dense and void regions -- i.e.\ they are cavitated -- and are mechanically unstable:\ $\partial P_{\rm IS}/\partial \rho \leq 0$).
The minimum of $P_{\rm IS}(\rho)$ at $\rho = \rho_S$ corresponds to an onset of mechanical instability under increasing tension that is the athermal limit of the stretched-liquid spinodal, i.e.\ the $T\to 0$ limit of the density $\rho_c(T)$ at which the free-energy barrier to cavitation vanishes.\cite{inprac}
Thus, by studying liquids' Sastry transitions, we can learn more about their cavitation under real-world conditions.
Altabet \textit{et al.}\ recently examined the Sastry transition of model glassforming liquids in much greater detail;\cite{altabet16,altabet18} see Section \ref{sec:ads}.
However, they did not study simple monatomic liquids, which readily crystallize, and in which cavitation and crystallization can complete.

In this paper, we examine the Sastry transition in monatomic liquids with a wide variety of interaction potentials.
We find that in liquids with short- and moderate-ranged attractive forces, most ambient conditions for which cavitation is likely to occur -- temperatures between the triple point and critical point, pressures between the stretched-liquid spinodal and atmospheric -- correspond to densities that are well below $\rho_S$.
In contrast,  liquids with long-ranged attractive forces, such as those formed by alkali metals, have  a broad region of thermodynamic phase space where cavitation is likely to occur and $\rho > \rho_S$.
Taken together, these results suggest that the barriers to cavitation in most (but not all) simple monatomic liquids under ambient conditions where significant cavitation is likely to occur arise primarily from the vibrational energy and entropy rather than the configurational energy of the EL basins they are most likely to occupy.
We also find that all liquids' IS are isostructural (have nearly identical local structure away from the voided regions) for $\rho < \rho_S$ but are isomorphic (exhibit a hidden scale invariance\cite{dyre14}) for $\rho > \rho_S$.

\section{Theoretical Background}
\label{sec:ads}

The Sastry transition can be better understood by placing it in context with CNT.
Cavitation bubbles in stretched model liquids are typically empty or nearly-empty.\cite{kuskin10}
Consider nucleation of an empty spherical bubble of radius $R$ in a liquid with density $\rho$ and free energy density $f$ at temperature $T$.
According to CNT, the free energy barrier to nucleation of this bubble is $\Delta F(\rho, T) = -(4\pi R^3/3)f(\rho,T) + 4\pi R^2 \gamma(\rho,T)$, where we allow the surface tension $\gamma$ to depend on $\rho$ and $T$ but not $R$.
Writing $f = u - Ts$, where $u$ and $s$ are the liquid's potential energy and entropy densities, and breaking $\Delta F(\rho,T)$ into its configurational and vibrational parts, one can write\cite{stillinger82,debenedetti01}
\begin{equation}
\displaystyle\frac{\Delta F(\rho,R,T)}{4\pi R^2} =   \Delta_{\rm conf}(\rho,R) +  \Delta_{\rm vib}(\rho,R,T) - T\Delta_{\rm ent}(\rho,R),
\label{eq:deltaF}
\end{equation}
where
\begin{equation}
\begin{array}{c}
\Delta_{\rm conf}(\rho,R) =  \gamma_{\rm conf}(\rho) - u_{\rm conf}(\rho) \displaystyle\frac{R}{3}   \\
\\
 \Delta_{\rm vib}(\rho,R,T) =   \gamma_{\rm vib}(\rho,T) - u_{\rm vib}(\rho,T) \displaystyle\frac{R}{3}  \\
\\
\Delta_{\rm ent}(\rho,R) =   \gamma_{\rm s}(\rho) - \left[ s_{\rm conf}(\rho) + s_{\rm vib}(\rho) \right]  \displaystyle\frac{R}{3} 
\end{array}.
\label{eq:barterms}
\end{equation}
Here $\gamma_s(\rho)$ is the entropic part of $\gamma$ arising from (e.g.) changes in the ordering of a liquid near a free surface.
Thus the free-energy barrier to cavitation has three components:\ configurational-energetic ($\Delta_{\rm conf}$), vibrational-energetic ($\Delta_{\rm vib}$), and entropic ($-T\Delta_{\rm ent}$).
According to Sastry \textit{et al.}'s picture,\cite{sastry97} when $\rho = \rho_S$ and $T=0$, the sum of the first two terms in Equation \ref{eq:deltaF} goes to zero in the limit $R \to 0$.
This explains why the Sastry transition can be interpreted as the $T\to 0$ limit of the stretched-liquid spinodal.\cite{inprac}
Liquids with $\rho < \rho_S$ can cavitate by proceeding directly down the same basin of their EL they currently occupy, whereas liquids with $\rho > \rho_S$ cannot;\ their cavitation must involve basin hopping.

Altabet, Stillinger and Debendetti recently showed\cite{altabet16}  that the system-size dependence of $P_{\rm IS}(\rho)$ is consistent with finite-size rounding of a first-order athermal phase transition.
The transition from cavitated to homogeneous IS at $\rho = \rho_S$ is a feature that is wiped out by the anharmonic intrabasin distortions that occur upon returning to the initial liquid thermodynamic state; this structure-obscuring behavior of intrabasin vibrational motion is consistent with the presence of a positive free energy barrier for cavitation in the liquid.\cite{still}
They also argued that in the thermodynamic limit, kinks in $f_{\rm conf}(\rho,T)$ and $f_{\rm vib}(\rho,T)$ at $\rho_S$ respectively produce discontinuous decreases and increases in $P_{\rm IS}(\rho)$ and $P_{\rm vib}(\rho,T)$ as $\rho$ exceeds $\rho_S$.
Since the liquid-state pressure $P_{\rm liq}(\rho,T) = P_{\rm IS}(\rho) + P_{\rm vib}(\rho,T) + \rho k_B T$ is continuous at $\rho_S$, these discontinuities must cancel.
In Ref.\ \cite{altabet18} they showed that these phenomena are not universal.
Strongly cohesive systems (i.e.\ systems with sufficiently deep and wide pair-potential wells) show the abovementioned behavior.
In these systems, the discontinuity in $P_{\rm IS}(\rho)$ is in fact associated with a first-order athermal phase transition between homogeneous and cavitated IS.
Weakly cohesive systems, while still possessing the Sastry minimum in $P_{\rm IS}(\rho)$, have a system-size-independent $P_{\rm IS}(\rho)$ that suggests no such first-order transition is present.

To the best of our knowledge, Refs.\ \cite{altabet16,altabet18} are the only two published particle-based simulation studies that have systematically examined how varying the pair interaction potential affects cavitation.
These studies employed Kob-Andersen\cite{kob95} and Wahnstr{\"o}m\cite{wahnstrom91} glass-forming binary mixtures of particles interacting via force-shifted versions of the ``n-6'' pair potential
\begin{equation}
U^A_{n-6}(r) = \displaystyle\frac{\epsilon}{n-6} \left[ 6\cdot 2^{n/6} \left( \frac{\sigma}{r} \right)^n - 2n \left( \frac{\sigma}{r} \right)^6 \right].
\label{eq:uAlt}
\end{equation}
Their force-shifting protocol was
\begin{equation}
\small
U_{fs}(r) = \bigg{ \{ } \begin{array}{ccc}
U^A_{n-6}(r) - U^{A}_{n-6}(r_c) - (r - r_c) U'(r_c)& , & r < r_c\\
0 & , & r > r_c
\end{array}.
\label{eq:ufs}
\end{equation}
Ref.\ \cite{altabet16} compared results for $n = 7$ and $n = 12$ systems with $r_c = 3.5\sigma$, and showed that differences between these systems were primarily quantitative rather than qualitative.
Ref.\ \cite{altabet18} compared results for $n = 7$ systems with various $1.4\sigma \leq r_c \leq 3.5\sigma$, and showed that systems with $r_c < 1.7\sigma$ ($r_c > 1.7\sigma$) are weakly (strongly) cohesive.
However, neither study isolated the effects of the attractive interactions' \textit{range} from the effects of their \textit{strength} (i.e.\ the depth of the pair potential well).

\section{Model and Methods}

One widely used generalization of the Lennard-Jones potential is the ``Mie'' potential
\begin{equation}
U_{\rm n}(r) = \epsilon \left[ \left( \displaystyle\frac{\sigma}{r} \right)^{2n} - 2\left( \displaystyle\frac{\sigma}{r} \right)^n \right].
\label{eq:un}
\end{equation}
Here $\epsilon$ and $\sigma$ are characteristic energy and length scales, and the exponent $n$ characterizes  the steepness of the repulsive and attractive interactions.
$U_n(r)$ is a general repulsive-attractive potential that can be used to model systems ranging from alkali metals ($n \simeq 4$) to noble gases ($n \simeq 6$) to colloids and buckyballs ($n \simeq 16$).\cite{clarke86,wales96,calvo12} 
Although the Morse potential $U_\alpha(r) = \epsilon \left( \exp[-2\alpha(r - \sigma)] - 2\exp[-\alpha(r - \sigma)] \right)$ is probably a more accurate model for some of these systems,\cite{wales96,calvo12} Lennard-Jones-type potentials are more widely used to model simple liquids.

\begin{figure}[h]
\includegraphics[width=3.25in]{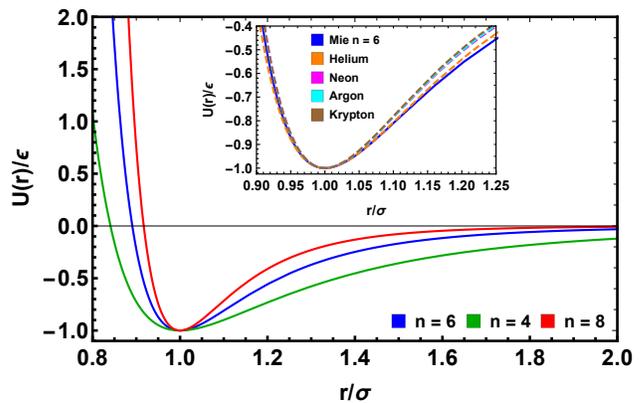}
\caption{Pair interaction potentials $U_n(r)$ for selected $n$.  For $n = 4$ multiple neighbor shells contribute to the thermodynamics of both solids and liquids.  For $n = 8$ the thermodynamics are dominated by the nearest neighbor shell.  
The inset shows a comparison of $U_6(r)$ to best-fit analytic potentials for noble gases obtained from many-body expansions of the interaction energies obtained from ab initio calculations.\cite{schwerdtfeger06}}
\label{fig:pot}
\end{figure}

Typical dynamical simulations employ a truncated-and-shifted version of $U_n(r)$.
Simulations including energy minimizations that find systems' IS, however, require modifying $U_n(r)$ in such a way that both $U_n(r)$ and $\partial dU_n/\partial r$ go to zero at some cutoff radius.\cite{wales04} 
This modification is typically achieved by multiplying $U_n(r)$ by some function $f$ that varies from $1$ to zero over the range $r_i \leq r \leq r_o$, where $r_i$ and $r_o$ are inner and outer cutoff radii.
We use the form of $f$ developed by Mei \textit{et al.}\cite{mei91}:
\begin{equation}
f(x) = \Bigg{ \{ } \begin{array}{ccc}
1 & , & x \leq 0 \\
(1-x)^3(1 + 3x + 6x^2) & , & 0 \leq x \leq 1 \\
0 & , & x > 1
\end{array},
\end{equation}
where $x = (r-r_i)/(r_o - r_i)$.  
We choose an $n$-dependent $r_i$ using the criterion $U_n[r_i(n)] = -\epsilon/100$, which yields $r_i(n) = [ 1 - \sqrt{99}/10 ]^{-1/n}\sigma$, and $r_o(n) = 11r_i(n)/10$.
Thus the interaction potential employed in this study is
\begin{equation}
U^{*}_n(r) = U_n(r)f\left[ \displaystyle\frac{r - r_i(n)}{r_o(n) - r_i(n)} \right].
\label{eq:unstar}
\end{equation}
$U^{*}_n(r)$ s plotted for selected $n$ in Figure \ref{fig:pot}, and values of $r_o(n)$ are given in Table \ref{tab:params}.
These choices ensure that any effects of imposing the smoothed cutoff [i.e.\ of the differences between $U^*_n(r)$ and $U_n(r)$] are small.

Comparing these potentials to best-fit analytic pair potentials for noble gases\cite{schwerdtfeger06,smits20} and metals\cite{wales96} makes it clear that the range of $n$ examined here is sufficiently wide to capture the behavior of all neutral monatomic liquids.
While quantitatively capturing the behavior of non-noble-gas elemental materials requires the use of $3$-body and higher-order interaction energies,\cite{schwerdtfeger06,li98} our focus here is on qualitative trends.

\begin{table}[h]
\caption{
Outer cutoff radii $r_o(n)$, nearest-neighbor distances $a(n)$, binding energies $E_{\rm FCC}(n)$ and densities  $\rho_{\rm FCC}(n)$ of the minimal-energy FCC lattices for the interaction potential $U^*_n(r)$ used in this study (Eq.\ \ref{eq:unstar}).  
Note that the $n\to\infty$ limits of these quantities are $r_o(\infty) = 1.1\sigma$, $a(\infty) = \sigma$, $E_{\rm FCC}(\infty) = -6\epsilon$, and $\rho_{\rm FCC}(\infty) = \sqrt{2}\sigma^{-3}$.}
\begin{ruledtabular}
\begin{tabular}{ccccc}
$n$ & $r_o(n)/\sigma$ & $a(n)/\sigma$ & $|E_{\rm FCC}(n)|/\epsilon$ & $\rho_{\rm FCC}(n)\sigma^3$ \\
4 & 4.1341 & .870031 & 17.93195 & 2.14739 \\
4.5 & 3.5685 & .917008 & 13.05351 & 1.83398 \\
5 & 3.1724 & .944147 & 10.53204 & 1.68034 \\
6 & 2.6590 & .970688 & 8.180256 & 1.54624 \\
7 & 2.3440 & .985107 & 7.22815 & 1.47933 \\
8 & 2.1325 & .992505 & 6.71965 & 1.44649 \\
\end{tabular}
\end{ruledtabular}
\label{tab:params}
\end{table}

Previous studies\cite{sastry98,debenedetti01} have shown that the character of the EL basins (i.e.\ inherent structures) preferentially sampled by liquids can change qualitatively as $T$ increases or decreases.
Hence, when comparing results for systems interacting via different pair potentials, it is helpful to define a dimensionless temperature $\tilde{T} = k_B T/E^*$, where $E^*$ is a system-dependent characteristic energy scale, and then compare systems at the same $\tilde{T}$.
The maximum basin depth for $N$-particle systems interacting via the pair potential $U^*_n(r)$ is $NE_{\rm FCC}(n)$, where $E_{\rm FCC}(n)$ is the binding energy of atoms in perfect FCC crystals at zero pressure and temperature (Table \ref{tab:params}). 
We set $E^*(n) = E_{\rm FCC}(n)/E_{\rm FCC}(6)$ and compare the inherent structures of liquids equilibrated at various temperatures in the range $0.5 \leq \tilde{T} \leq 1.5$, i.e.\ $E_{\rm FCC}(n)/2E_{\rm FCC}(6) \leq k_B T \leq 3E_{FCC}/2E_{\rm FCC}(6)$.
We chose this $E^*(n)$ to facilitate comparison of our results to the extensive literature on cavitation in standard Lennard-Jones ($n = 6$) systems:\ for these systems, $0.5 \leq \tilde{T} \leq 1.5$ corresponds to the temperature range $0.5 \leq k_B T/\epsilon \leq 1.5$.

We generate equilibrated liquids with a wide range of densities and their IS using standard molecular dynamics and energy-minimization techniques.
$N = 4000$ atoms, each of mass $m$, are placed in cubic simulation cells, and periodic boundary conditions are applied along all three directions.
Temperature is maintained using a Langevin thermostat.
After thorough equilibration, $NVT$ runs are continued while periodic snapshots of the liquids' configurations are taken.\cite{foottimestep}
These snapshots are then energy-minimized using the Polak-Ribi{\'e}re conjugate-gradient algorithm\cite{polak69} to find the liquids' IS. 
All simulations are performed using LAMMPS.\cite{plimpton95}

We will compare systems' Sastry densities $\rho_S(n, \tilde{T})$ to their spinodal vaporization densities $\rho_v(n,\tilde{T})$ and their equilibrium crystallization densities $\rho_x(n,\tilde{T})$.
We estimate $\rho_v(n,\tilde{T})$ as the density for which liquids' $P(\rho,\tilde{T})$ are minimized, i.e.\ the density below which the liquid phase is mechanically unstable. 
No $\rho_v$ values are reported for ($n,\tilde{T}$) that lack clear minima; estimating these systems' vaporization densities is more difficult,\cite{mastny07,toxvaerd15}  and is not essential here.
We estimate $\rho_x(n,\tilde{T})$ using the Stevens-Robbins protocol.\cite{stevens93}
Specifically, we start from a perfect FCC crystal at the given $\rho$ and equilibrate it at the given $\tilde{T}$ for $400\tau$.
We then freeze half the system in place while equilibrating the other half at  $3\tilde{T}/2$ for another $400\tau$ to create coexisting liquid and crystalline regions within the simulation cell.
Finally, we unfreeze the crystalline half, reset the liquid half's kinetic temperature to $\tilde{T}$, and integrate the system forward in time for at least another $10^4\tau$.
Crystallization occurs over this period if $\rho \geq \rho_x(n,\tilde{T})$.
Strictly speaking, $\rho_x(n,\tilde{T})$ is the density at which the Helmholtz free energies of the liquid and crystalline phases are equal.
Note that the $\rho_v(n,\tilde{T})$ and $\rho_x(n,\tilde{T})$ obtained using these methods are rather sensitive to both system size and the choice of potential cutoff.\cite{mastny07,toxvaerd15}  
However, our focus in this paper is on qualitative trends, and none of the results presented below would be altered by small changes in $\rho_v$ or $\rho_x$.

\section{Results}
\label{sec:results}

\begin{figure}[h]
\includegraphics[width=3.25in]{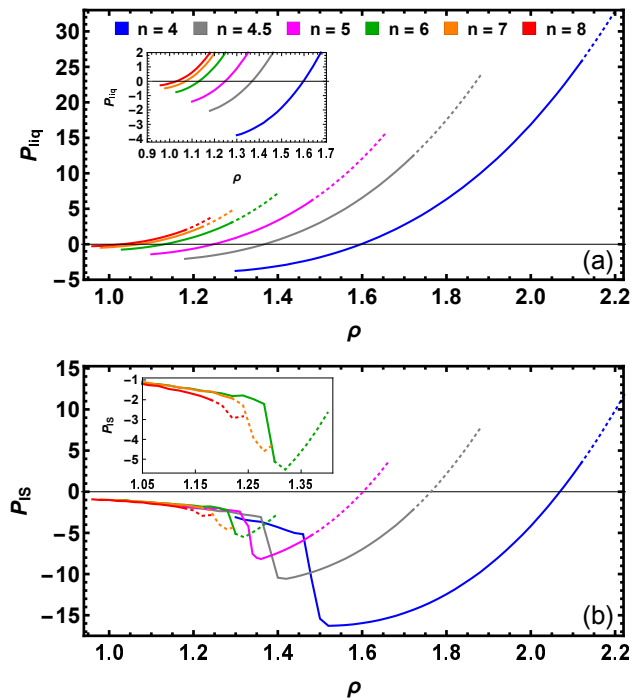}
\caption{Equations of state for $\tilde{T} = .75$ liquids and their inherent structures.  Solid curves in panels (a-b) respectively show $P_{\rm liq}(\rho)$ and  $P_{\rm IS}(\rho)$ for $\rho_v(n,\tilde{T}) \leq \rho \leq \rho_x(n,\tilde{T})$, while dotted curves show these quantities in metastable liquids with $\rho > \rho_x(n,\tilde{T})$.   The inset to panel (a) shows a zoomed-in view of the same data, and the inset to panel (b) is a zoomed-in version of the $6 \leq n \leq 8$ data that highlights how $\rho_S > \rho_x$ for these systems.}
\label{fig:PliqPIS}
\end{figure}

\begin{figure*}[t]
\includegraphics[width=7in]{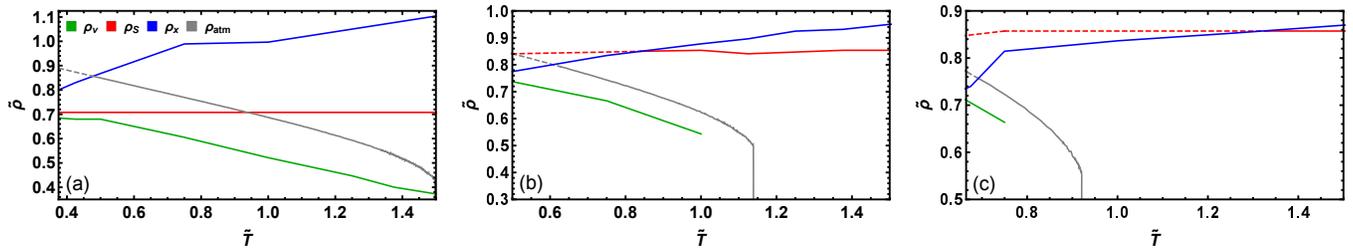}
\caption{Phase diagrams for [panel (a)] $n = 4$, [panel (b)] $n = 6$, and [panel (c)] $n = 8$. Here $\tilde{\rho} = \rho/\rho_{\rm FCC}(n)$.  Dashed curves indicate $\rho_S$ and $\rho_{\rm atm}$ for metastable liquids with $\rho < \rho_x(\tilde{T})$,}
\label{fig:phase}
\end{figure*}

\begin{figure}[h]
\includegraphics[width=3.25in]{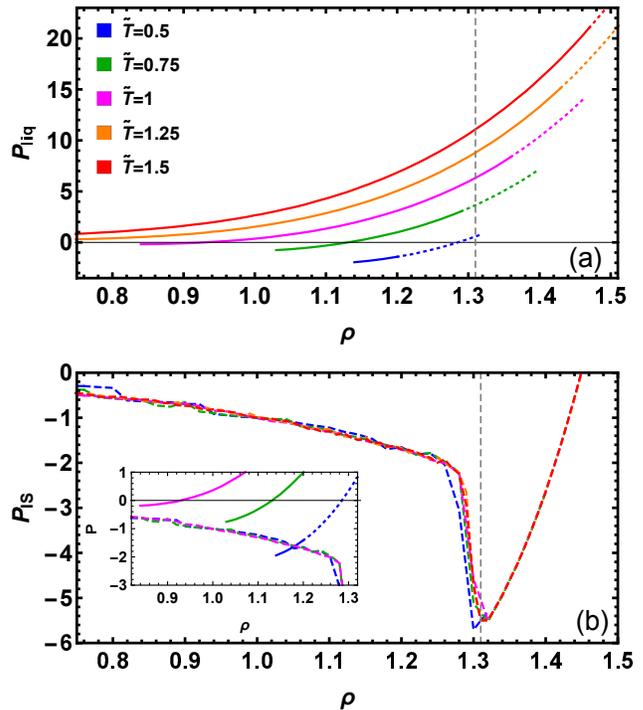}
\caption{Equations of state for $n=6$ (Lennard-Jones) liquids and their inherent structures. Solid curves show $P_{\rm liq}(\rho)$ for  $\rho_v(\tilde{T}) \leq \rho \leq \rho_x(\tilde{T})$, dotted curves show  $P_{\rm liq}(\rho)$ in metastable liquids with $\rho > \rho_x(\tilde{T})$, and dashed curves show $P_{\rm IS}(\rho)$.  The vertical dashed line indicates the mean $\rho_S = 1.31\sigma^{-3}$.}
\label{fig:PliqPISn6}
\end{figure}

\begin{figure}[h]
\includegraphics[width=3.05in]{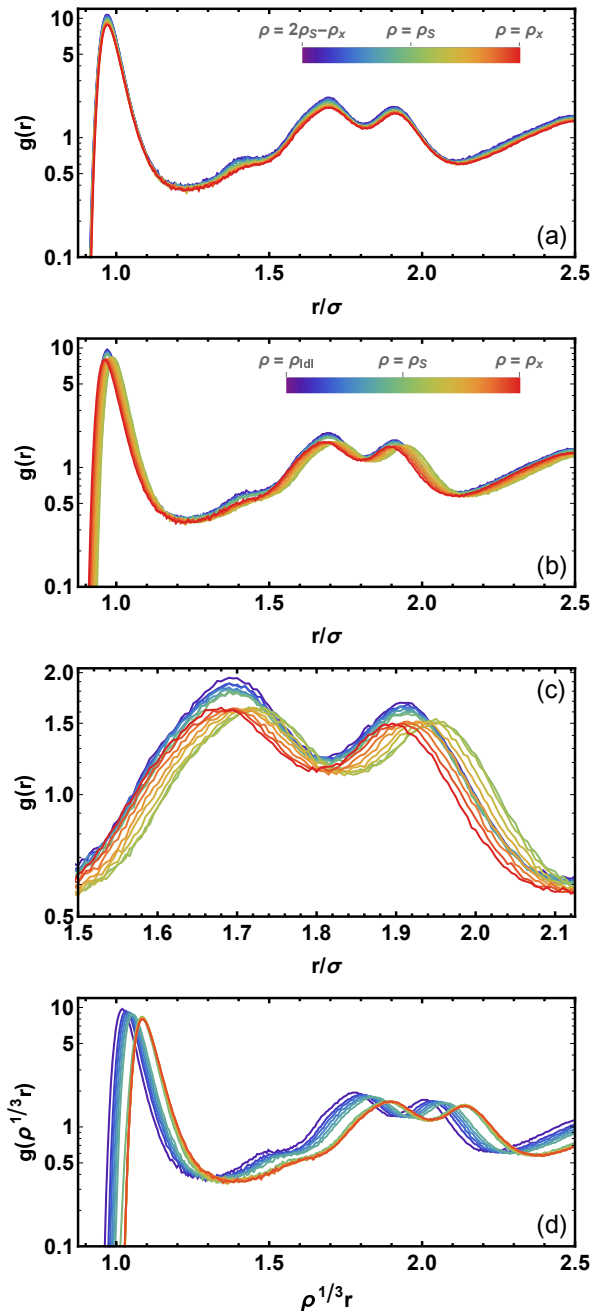}
\caption{Pair correlations in IS of $n = 6$ systems af [panel (a)] $\tilde{T} = 0.75$ and  [panels (b-d)] $\tilde{T} = 1.375$.   Panel (c) is a zoomed-in version of panel (b).  Here $\rho_{\rm ldl} = 4\rho_x(6,1.375)/5$.  Panel (d) shows the scaled pair correlation function $g(\rho^{1/3}r)$.  For $\rho < \rho_S$, lower density systems have slightly larger $g(r)$ for small $r$ because these systems are inhomogeneous.}
\label{fig:gofr}
\end{figure}

We begin by presenting results $4 \leq n \leq 8$ Mie liquids' and their inherent structures' equations of state.
All data in Figs.\ \ref{fig:PliqPIS}-\ref{fig:PliqPISn6}, \ref{fig:gofr} are averaged over 25 statistically independent samples.
 All densities discussed below are in units of $\sigma^{-3}$ and all pressures are in units of $\epsilon\sigma^{-3}$.

Figure \ref{fig:PliqPIS}(a) shows all systems' average liquid-state pressures $P_{\rm liq}(\rho)$ for $\tilde{T}=.75$.
For this $\tilde{T}$, all $4 \leq n \leq 8$ have clearly observable minima in $P(\rho)$ and hence clearly defined $\rho_v$.
Furthermore, all $n$ have $P(\rho_v) < 0$ and hence can be prepared as metastable ``stretched'' liquids.
The basic features of the data shown here are all expected.
Longer-range attractions allow liquids to sustain much larger tensile stress, and the narrowing of the range of densities and pressures for which liquids are at least metastable as $n$ increases is consistent with narrowing of the pair potential well.
Since these liquids span a very wide range of densities and pressures, we conclude that $\tilde{T} = .75$ is a good value with which to begin our detailed analyses.

Figure \ref{fig:PliqPIS}(b) shows these systems' average $P_{\rm IS}(\rho)$.
The trends shown here are qualitatively consistent with those reported in Refs.\ \cite{sastry97,altabet16,altabet18}.
The Sastry densities $\rho_S$ and pressures $P_S = P_{\rm IS}(\rho_S)$ respectively increase and decrease with increasing $n$ as the pair interactions soften and attractive forces become longer-ranged.
The kinks in $P_{\rm IS}(\rho)$ at $\rho = \rho_{\rm o}$ and $\rho = \rho_S$ both become more dramatic with decreasing $n$; here $\rho_{\rm o} < \rho_S$ is the density below which all IS are cavitated and at which $\partial P_{\rm IS}/\partial \rho$ drops sharply.\cite{altabet16}
While $P_{\rm IS}(\rho)$ has large finite-$N$ corrections for strongly-cohesive (e.g.\ low-$n$) systems,\cite{altabet16,altabet18} and we do not attempt to address finite-system-size-related issues in this paper, one should remain aware that the $N$- and $n$-dependences of the phenomena we discuss below are likely coupled.

Figure \ref{fig:PliqPIS} also illustrates a phenomenon which has not been previously discussed:\ the interaction of the Sastry and crystallization transitions.
Dotted curves indicate results for metastable liquids with $\rho > \rho_x$.
For long-ranged potentials ($n < 6$), $\rho_S$ is well below $\rho_x$ for this $\tilde{T}$.
For Lennard-Jones and shorter-ranged potentials, however, $\rho_S > \rho_x$ -- increasingly so as $n$ increases.
If $\rho_S > \rho_x$ for a given $\tilde{T}$, the entire range of $\rho$ for which liquids are thermodynamically stable at that $\tilde{T}$ lies below the Sastry density.
Dynamic cavitation of these liquids under further hydrostatic expansion seems unlikely to be governed directly by the Sastry transition.

\begin{table}[h]
\caption{Mean liquid-state pressures at the Sastry density:\ $\langle P_{\rm liq}(\rho_S,\tilde{T}) \rangle \sigma^3/\epsilon$.  Values for metastable supercooled liquids (systems with $\rho_S > \rho_x$) are italicized.  The temperatures $\tilde{T}_0(n)$ for which $P_{\rm liq}(\rho_S,\tilde{T}) = 0$ are given in Table \ref{tab:Tstar0}.}
\begin{ruledtabular}
\begin{tabular}{cccccc}
$\tilde{T}$ & $n = 4$ & $n = 5$ & $n = 6$ & $n = 7$ & $n = 8$\\ 
0.5 & -5.44 & \it{-1.04} & \it{0.336} & \it{--1.04} & \it{-1.06} \\
0.75 & -1.51 & 1.99 & \it{3.65} & \it{4.32} & \it{3.72} \\
1.0 & 1.97 & 4.64 & 6.69 & \it{7.11} & \it{6.54} \\
1.25 & 4.12 & 7.42 & 7.45 & 9.72 & 9.12 \\
1.5 & 8.15 & 9.32 & 11.6 & 12.1 & 13.8 
\end{tabular}
\end{ruledtabular}
\label{tab:meanliqPsast}
\end{table}

Another important question in determining the Sastry transition's relevance for a given system is:\ what is $P_{\rm liq}(\rho_S,T)$?
If $P_{\rm liq}(\rho_S,T)$ is positive and well above the liquid's vapor pressure $P_{\rm vap}(\rho_S,T)$, cavitation of liquids with $\rho = \rho_S$ and temperature $T$ is highly unlikely.
On the other hand, if $P_{\rm liq}(\rho_S,T) \lesssim P_{\rm vap}(\rho_S,T)$, cavitation is far likelier.
For these systems, the energy barriers for cavitation at $\rho = \rho_S - \delta\rho$ should be signifcantly smaller than those for cavitation at $\rho = \rho_S + \delta\rho$ is (where $0 < \delta\rho \ll \rho_S$),\cite{altabet16} so cavitation is likely to be tightly coupled to -- and effectively nucleated by -- local density or temperature fluctuations within the liquid.\cite{sastry97,wang09}

Table \ref{tab:meanliqPsast} lists values of $P_{\rm liq}(\rho_S,\tilde{T})$ for selected $n$ and $\tilde{T}$.
$P_{\rm liq}(\rho_S, 0.5)$ is small or negative for all $n$; the Sastry transition is likely to strongly influence cavitation in these liquids.
However, $\tilde{T}=0.5$ liquids are metastable for $n \geq 5$. 
In these systems, local density fluctuations to $\rho = \rho_S + \delta\rho$ are likely to nucleate crystallization \textit{and} local density fluctuations to $\rho = \rho_S - \delta\rho$ are likely to nucleate cavitation; this competition gives rise to very complicated physics.\cite{baidakov11}
As $\tilde{T}$ increases, $P(\rho_S)$ increases rapidly, but remains negative or small for a wider range of $\tilde{T}$ in systems with longer-ranged attractions.
For $\tilde{T} \gtrsim 1$, however, $P_{\rm liq}(\rho_S)$ is well above $P_{\rm vap}(\rho_S)$ for all $n$.
It seems unlikely that the Sastry transition has much influence on these liquids' quiescent-state mechanical properties, but it may still strongly influence their cavitation under dynamic hydrostatic expansion.

The above results suggest that examining how $\rho_S(n,\tilde{T})$ compares to $\rho_v(n,\tilde{T})$ and $\rho_x(n,\tilde{T})$ for a wide range of $n$ and $\tilde{T}$ can shed a great deal of light on how the Sastry transition influences cavitation in liquids with a wide range of pair interactions.
Figure \ref{fig:phase} shows phase diagrams for $n = 4$, $6$, and $8$.
For perspective, all plots also show $\rho_{\rm atm}(\tilde{T})$, the equilibrium density at the ``atmospheric'' pressure $P_{\rm atm} = .01k_BT/\sigma^3$.
Here $\rho_{\rm atm}$ and $P_{\rm atm}$ are not rigorous quantities.
A more thorough study would replace $\rho_{\rm atm}(n,\tilde{T})$ with $\rho_{vl}(n,\tilde{T})$, the density of a Mie liquid that is in equilibrium with its own vapor, but calculations of this quantity are highly sensitive to $N$ and $r_o$\cite{mastny07,toxvaerd15} and are beyond our present scope.
Nevertheless, the  $\rho_{\rm atm}(\tilde{T})$ curves and their discontinuous drops at the boiling points $\tilde{T}_{\rm boil}(n)$\cite{supboil} provide context for what we will describe below.

As expected,\cite{sastry97} $\rho_S$ is independent of $\tilde{T}$ to within the accuracy of our measurements, for all $n$.
Otherwise the topology of these phase diagrams depend strongly on $n$.
To further clarify how these topologies relate to cavitation, we define four characteristic regions of thermodynamic phase space. 
Region 1 consists of all ($\rho,T$) for which $\rho_S < \rho < \rho_x$ and $\rho > \rho_{\rm atm}$.
Region 2 consists of all ($\rho,T$) for which $\rho_S < \rho < \rho_x$ and $\rho < \rho_{\rm atm}$.
Region 3 consists of all ($\rho,T$) for which $\rho_v < \rho < \rho_S$ and $\rho < \rho_{\rm atm}$.
Finally, region 4 consists of all ($\rho,T$) for which $\rho_v < \rho < \rho_S$ and $\rho > \rho_{\rm atm}$.
Cavitation is least likely in region 1 and most likely in region 3.
The clearest-cut scenario for the Sastry transition to play the dominant role in controlling cavitation is dynamic hydrostatic expansion from regions 1 or 2 into region 3.
It may also strongly influence quiescent cavitation in the lower portions of region 2 and upper portions of region 3.

Next we define two characteristic temperatures for these systems.  
 $\tilde{T}^*$ is the temperature at which $\rho_S = \rho_x$.  
 For all $\tilde{T} < \tilde{T}^*$, $\rho_S > \rho_x$; the Sastry transition lies in a region of phase space where (for the given $\tilde{T}$) the thermodynamically stable phase is crystalline.
$\tilde{T}^*$ is also the lower boundary of region 1.
$\tilde{T}_0$ is the temperature at which $P_{\rm liq}(\rho_S, \tilde{T}) = 0$.
For $\rho \simeq \rho_S$ and $\tilde{T}$ well above $\tilde{T}_0$, cavitation is unlikely because the thermal barriers to it ($\Delta_{\rm vib}$ and $-T\Delta_{\rm ent}$) are large.
Values of  $\tilde{T}^*$ and $\tilde{T}_0$ for all systems are given in Table \ref{tab:Tstar0}.

\begin{table}[h]
\caption{Characteristic reduced temperatures for Mie liquids. ``$--$'' indicates that $\tilde{T}^*$ is below our lowest simulated value ($0.375$ for $n = 4$).}
\begin{ruledtabular}
\begin{tabular}{cccccc}
Quantity & $n = 4$ & $n = 5$ & $n = 6$ & $n = 7$ & $n = 8$\\ 
$\tilde{T}^*$ & $--$ & $0.51$ & 0.84 & 1.20 & 1.33 \\
$\tilde{T}_0$ & 0.85 & 0.60 & 0.47 & 0.68 & 0.64 \\
\end{tabular}
\end{ruledtabular}
\label{tab:Tstar0}
\end{table}

For $n = 8$ liquids, $\tilde{T}^*$ is well above $T_{\rm boil}$\cite{supboil} and almost certainly above the critical temperature $\tilde{T}_{\rm crit}$.\cite{cailol98}
For $n = 7$ liquids $\tilde{T}^* \simeq T_{\rm boil}$.\cite{supboil} 
Thus it seems unlikely that the Sastry transition heavily influences cavitation in (initially) thermodynamically-stable liquids with short-ranged pair interactions.
This result is consistent with the weak, broad minima in these systems' $P_{\rm IS}(\rho)$ shown in Fig.\ \ref{fig:PliqPIS}. 
Overall these liquids' behavior is similar to that of Altabet \textit{et al.}'s weakly cohesive liquids.\cite{altabet18}

All thermodynamically stable $n \simeq 6$ liquids lack region 2.
Dynamic hydrostatic expansion of $n \simeq 6$ liquids that are initially in region 1 passes through the upper (high-pressure) regions of region 4, and reaches region 3 only for densities that are well below $\rho_S$.
Thus it also seems unlikely that the Sastry transition heavily influences cavitation in (initially) thermodynamically-stable monatomic $n \simeq 6$ liquids such as noble liquids (Fig.\ \ref{fig:pot}).
However, this result also indicates that the free-energy barriers to cavitation in these systems are primarily vibrational-energetic and entropic rather than configurational-energetic, i.e. they are dominated by y $\Delta_{\rm vib}$ and $-T\Delta_{\rm ent}$.
Moreover, the Sastry transition may indeed play a major role in \textit{supercooled} $n \simeq 6$ liquids, e.g.\ Lennard-Jones liquids with $\tilde{T} \lesssim 0.6$.
As discussed above, these liquids occupy a large region of thermodynamic phase space; note that Lennard-Jones liquids that are metastable with respect to both crystallization and cavitation can be prepared for $\tilde{T}$ as low as $0.35$.\cite{baidakov11,baidakov14}

For $n = 4$, regions 1 and 2 are both very large.
There is a wide range of thermodynamic  phase space where dynamic hydrostatic expansion can take these liquids from region 1 though region 2 into region 3, or directly from region 2 into region 3.
We expect that the Sastry transition is likely to play a crucial role in these systems, for both hydrostatic-expansion-driven cavitation and density-fluctuation-driven cavitation (particularly in region 2).
For $\tilde{T} < 0.85$, these liquids have negative pressure at $\rho = \rho_S$.
Thus Sastry and Altabet \textit{et al.}'s picture suggests that $n=4$ stretched liquids are (meta)stabilized against cavitation by both vibrational-energetic and entropic barriers ($\Delta_{\rm vib}$ and $-T\Delta_{\rm ent}$, respectively).

For the remainder of this Section, we will focus on $n = 6$ (Lennard-Jones) systems since they best capture the physics of the majority of real atomic liquids.  
Figure \ref{fig:PliqPISn6} shows the equations of state for $n = 6$ liquids and their IS for a wide range of $\tilde{T}$.
Panel (a) shows that for our choice of system size and cutoff radius, metastable stretched liquids can be prepared for $\tilde{T} \lesssim 1$. 
It illustrates how liquids with $\rho = \rho_S$ are metastable (supercooled) for $\tilde{T} < \tilde{T}^*$, and also how the range of densities and pressures over which these liquids are stable increases rapidly with increasing $\tilde{T}$.
The Sastry transition is most likely to be relevant when these liquids are isochorically cooled (for $\rho < \rho_S$) or dynamically hydrostatically expanded (from an initial $\rho > \rho_S$).

Panel (b) illustrates how these liquids' IS'  equations of state depend on $\tilde{T}$.
Remarkably, the lowest-density region of the $\tilde{T} = 0.5$ liquid's equation of state has $P_{\rm liq}(\rho,T) < P_{\rm IS}(\rho)$.
The identity $P_{\rm liq}(\rho,T) \equiv P_{\rm IS}(\rho) + P_{\rm vib}(\rho,\tilde{T}) + \rho k_B T$\cite{stillinger99} implies that these liquids have $P_{\rm vib}(\rho,\tilde{T}) \leq -\rho k_B T$, i.e.\ that thermal vibrations in these liquids substantially reduce their pressure.
Since $P_{\rm vib}(\rho) = 0$ at $T = 0$, we know $(\partial P_{\rm vib}/\partial T)_\rho$ is negative for some $0 \leq T \leq .5\epsilon/k_B$ for these systems.
Since  $(\partial P_{\rm IS}/\partial T)_\rho \equiv 0$ for small $T$, and given the Maxwell identity $(\partial P/\partial T)_\rho  = \alpha_{\rm th}/\kappa$ [where $\alpha_{\rm th}$ is the thermal expansion coefficient and $\kappa$ is the compressibility], it must be this negative  $\partial P_{\rm vib}(\rho)/\partial T$ that leads to the thermodynamic instability encountered when $\kappa$ becomes negative.
This signature of systems that cavitate when isochorically cooled has not (to our knowledge) been previously reported; it may vanish as $N$ increases.

\begin{figure}[h]
\includegraphics[width=3.375in]{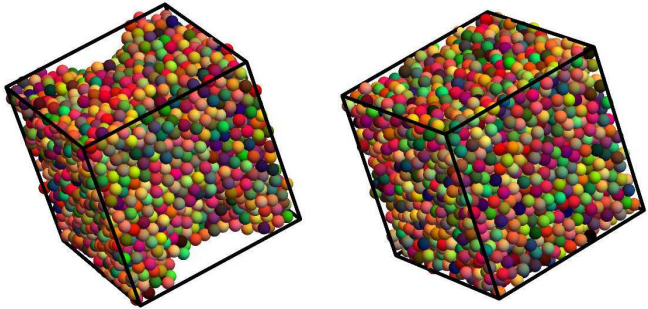}
\caption{Snapshots of typical inherent structures for (left panel) $\rho = .97\rho_S$ and (right panel) $\rho = 1.03\rho_S$, for $\tilde{T}=1.375$.  The snapshots are plotted at a common scale.}
\label{fig:snapshots}
\end{figure}

We conclude our analyses by switching our focus from the Sastry transition's macroscopic features to its microscopic features.
Figure \ref{fig:snapshots} shows snapshots for typical $n = 6$, $\tilde{T} = 1.375$ inherent structures for densities slightly below and above $\rho_S$.
As reported in Refs.\ \cite{sastry97,altabet16,altabet18}, IS are inhomogeneous and cavitated for $\rho < \rho_S$ but homogeneous for $\rho > \rho_S$.
The left-hand snapshot shows an IS containing a single large void.
It clearly has a significant degree of locally-close-packed-crystalline order.
This cannot arise in the bidisperse mixtures employed in Refs.\ \cite{altabet16,altabet18}, which were designed\cite{kob95,wahnstrom91} to suppress crystallization.\cite{whySast}
The right-hand snapshot shows a homogeneous IS.
It also shows some signs of crystalline order, but the degree of ordering present is unclear.

Examining the IS' structure in greater detail yields additional insights.
Figure \ref{fig:gofr} shows their pair correlation functions $g(r)$ for $\tilde{T} = .75$ and $\tilde{T} = 1.375$.
For $\tilde{T} = .75$ [panel (a)], all systems have peaks in $g(r)$ at  $r \simeq a_n$, $r \simeq \sqrt{3}a_n$, and $r \simeq 2a_n$, where $a_n = [\sqrt{2}/\rho_{\rm FCC}(n)]^{1/3}$ is the equilibrium nearest-neighbor distance in the ground-state crystals (Tab.\ \ref{tab:params}).
These distances are characteristic of random-close-packed (RCP) order.\cite{torquato00}
The lower-density systems show a small additional peak at $r \simeq \sqrt{2}a_n$, which is the second-nearest neighbor distance in FCC crystals.
Overall, the results indicate that these liquids' IS are \textit{isostructural}:\  except for the void surfaces, they have the same density and $k$th-nearest-neighbor distances as the ground-state crystal.

Data for $\tilde{T} = 1.375$ [panels (b-c)] appear similar at first glance but are critically different in one respect. 
For $\rho < \rho_S$, the pattern is the same as for $\tilde{T} = .75$; $g(r)$ has peaks at $r \simeq a_n$, $r \simeq \sqrt{3}a_n$, and $r \simeq 2a_n$.
The height of these peaks increases slowly with decreasing $\rho$ because IS for $\rho < \rho_S$ are inhomogeneous.
At higher densities, however, the pattern is very different; the heights of the peaks are  $\rho$-independent, but their positions are $\rho$-dependent.
Specifically, the second- and third-nearest-neighbor distances decrease with increasing $\rho$.

Liquids at different ($\rho,T$) whose pair correlation functions collapse under the scaling $r \to \rho^{1/3}r$, i.e.\ have the same $g(\rho^{1/3}r)$, are ``isomorphic''.
Dyre \textit{et al.}\ have shown these ``Roskilde simple'' liquids have the same pressure-energy correlations, dynamics, and excess entropy, as well as  the same equation of motion in the reduced coordinates $\rho^{1/3}\vec{r}^N$.\cite{gnan12,dyre14,dyre16}
Isomorphism implies that the $k$th-nearest-neighbor distances scale with the typical intermonomer distance $a_\rho = \rho^{-1/3}$.
If, on the other hand, the $k$th-nearest-neighbor distances are $\rho$-independent, the $g(r)$ curves will collapse but the scaled $g(\rho^{1/3}r)$ curves will not.
Figure \ref{fig:gofr}(d) shows the scaled pair correlation functions $g(\rho^{1/3}r)$ for our $\tilde{T} = 1.375$ systems.
Clearly they collapse for $\rho > \rho_S$.
Here we have highlighted results for $\tilde{T} = 1.375$ in Figs.\ \ref{fig:snapshots}-\ref{fig:gofr} because the wider range of densities with $\rho_S < \rho < \rho_x$ for this $\tilde{T}$ allowed the nature of the collapse to be clarified, but similar collapses occur for other $\tilde{T}$.  
In particular, they also occur for $\tilde{T} = 1$ and $1.125$, i.e.\ they also occur for temperatures below the atmospheric-pressure boiling point.
Analogous results hold for other $n$.

Thus we have shown that Mie-liquid IS for $\rho > \rho_S(n,\tilde{T})$ are \textit{isomorphic}.
They possess essentially the same order; the main $\rho$-dependence of this order is that the characteristic $k$th-nearest-neighbor distances are proportional to $a_\rho  = \rho^{-1/3}$.
``Isomorphs'', curves in ($\rho,T$) phase space along which a given system is isomorphic, play a very important role in liquid-state physics; for example, the freezing and melting lines of Roskilde-simple liquids are isomorphs.\cite{dyre14}
The fact that $\rho > \rho_S$ IS are isomorphic suggests that the IS' equation of state is also an isomorph, albeit one of a different character given that it is a curve $\rho(P)$ rather than a curve $\rho(T)$ as is the case for freezing and melting lines.
More generally, the sharp contrast between $\rho < \rho_S$ isostructural IS and  $\rho > \rho_S$ isomorphic IS is further evidence that the Sastry transition is a useful concept for improving our fundamental understanding of cavitation.

\section{Discussion and Conclusions}

In this paper, we studied the Sastry transition in monatomic Mie liquids.
We showed that for short-ranged pair interactions ($n \gtrsim 7$), thermodynamically stable liquids with $\rho > \rho_S$ exist only at reduced temperatures $\tilde{T}$ where pressures are high, making cavitation unlikely.
Indeed, the minimum $\tilde{T}$ for which such liquids are found are above the liquids' ``atmospheric''-pressure boiling temperatures, and probably above their critical temperatures.
The Sastry transition is unlikely to a play a major role in these ``weakly cohesive''\cite{altabet18} liquids' physics.

Thermodynamically stable liquids with small or negative pressures and $\rho > \rho_S$ exist only when the pair interactions are long-ranged  ($n \lesssim 5$).
In these systems, local density fluctuations to $\rho = \rho_S - \delta$ are likely to significantly enhance cavitation.
More generally, the Sastry transition likely plays a crucial role in these systems' cavitation under dynamic hydrostatic expansion.
A prominent group of elements with such long-ranged pair interactions is the alkali metals.
The $n = 4$ Mie potential studied here and the Morse potential with $\alpha \simeq 3.2$ model such metals only roughly,\cite{clarke86,wales96} but followup studies using realistic many-body potentials\cite{li98} could shed light on the nature of cavitation in these systems.
Such studies could be especially useful because CNT is particularly inaccurate for systems with long-ranged attractions.\cite{oxtoby88}

All thermodynamically stable Lennard-Jones ($n=6$) liquids with $\tilde{T} \lesssim 0.84$ have $\rho < \rho_S$, suggesting that the Sastry transition is unlikely to heavily influence their cavitation under dynamic hydrostatic expansion.
The majority of previous simulation studies of the cavitation of these liquids\cite{sastry97,wang09,baidakov11,baidakov14,baidakov16,angelil14,kinjo98,baidakov05,kuskin10,cai16} have explored this regime.
However, this result also indicates the free-energy barriers to cavitation in these liquids are vibrational-energetic and entropic rather than configurational-energetic, i.e.\ that the barriers are dominated by the $\Delta_{\rm vib} - T\Delta_{\rm ent}$ term in Eq.\ \ref{eq:deltaF}.
This raises a fundamental question:\ are $\rho < \rho_S$ liquids stabilized (or metastabilized) against cavitation primarily by $\Delta_{\rm vib}$ or $-T\Delta_{\rm ent}$? 
$\Delta_{\rm ent}$ is not easy to calculate and hence the ratio $|\Delta_{\rm vib}/T\Delta_{\rm ent}|$ has been little studied, but it \textit{can} be calculated using state-of-the art methods;\cite{menzi16} these calculations have shown that the entropic term can be important even at moderate temperatures.
It would be very interesting to apply such methods in simulations of noble liquids that use accurate atomistic potentials obtained from many-body expansions of the interaction energies obtained from ab initio theories.\cite{schwerdtfeger06,smits20} 
Note that while CNT predictions of cavitation rates are notoriously inaccurate, much of this inaccuracy results from the failure of approximations CNT typically makes, such as the assumptions that cavities are uniform and have the same properties as the bulk gas, interfaces are atomically thin, and that $\gamma$ does not depend on $R$ or $T$.\cite{oxtoby88}

For $\tilde{T} \gtrsim 0.84$, we showed that thermodynamically stable Lennard-Jones liquids with $\rho > \rho_S$ exist, but the ambient pressures at $\rho = \rho_S$ are high, indicating that cavitation at these temperatures is likely only for densities well below $\rho_S$.
However, we also found that metastable supercooled $n \lesssim 6$ liquids with $\rho \gtrsim \rho_S$ and low ambient pressures $P \lesssim P_{\rm atm}$ occupy a broad region of thermodynamic phase space, at least for the small system size considered here.
In such systems, cavitation competes with crystallization, giving rise to complicated physics\cite{baidakov11} that is beyond our present scope.
Again, it would be very interesting to further examine the degree to which the Sastry transition influences cavitation in this regime, using either coarse-grained or more realistic models.
Our results suggest that it plays a major role.

The Sastry transition is well-known to correspond to the transition of IS' macroscopic structure from cavitated to homogeneous.
By studying model monatomic liquids, we found that it also corresponds to a sharp transition in IS' \textit{microscopic} structure.\cite{whySast}
IS for $\rho < \rho_S$ are isostructural:\ they have locally RCP order, with $k$th-nearest-neighbor distances that are $\rho$-independent.
In contrast, IS for $\rho > \rho_S$ are isomorphic:\ they also have locally RCP order, but with $k$th-nearest-neighbor distances that scale with the characteristic interparticle separation $a_\rho \equiv \rho^{-1/3}$.
The fact that this result holds for all $n$ we studied suggests that it may hold for all Roskilde-simple liquids.
Given the degree to which the physics of such liquids is universal and the wide variety of pair interactions that meet the criteria for  Roskilde-simplicity,\cite{schroder14} we claim that it should hold for most monatomic liquids that lack strongly directional interactions.
At the very least, this isostructural-isomorphic dichotomy may be useful in developing novel EL-based microscopic theories of cavitation.

The data that support the findings of this study are available from the corresponding author upon request.
We thank Frank H.\ Stillinger for helpful discussions.


\begin{thebibliography}{52}
\bibitem [{inp()}]{inprac}%
  \BibitemOpen
  \href@noop {} {}\bibinfo {note} {In practice, the stretched-liquid spinodal
  $\rho_c(T)$ is interrupted by crystallization or glass formation as $T$
  decreases.\cite{sastry00}}\BibitemShut {NoStop}%
\bibitem [{foo()}]{foottimestep}%
  \BibitemOpen
  \href@noop {} {}\bibinfo {note} {The varying steepness of the potentials
  (Fig.\ \ref{fig:pot}) requires an $n$-dependent MD timestep $dt(n)$ for
  maximum efficiency. For $\tilde{T} \leq 1$, we choose the criterion $dt(n) =
  6\sqrt{2}/[125\omega(n)]$, where $\omega(n) = \frac{\sigma}{\epsilon \tau}
  \sqrt{\frac{\partial^2 U_n(r)}{\partial r^2} |_{r = \sigma} } =
  \frac{\sqrt{2}n}{\tau}$ is the characteristic vibration frequency, and $\tau
  = \sqrt{m\sigma^2/\epsilon}$ is the Lennard-Jones time unit. The prefactor
  $6\sqrt{2}/125$ sets $dt(n) = 6\tau/(125n)$ [e.g.\ $dt(6) = .008\tau$], which
  is sufficiently small to obtain converged results for all $n$. For $\tilde{T}
  > 1$ we use slightly smaller timesteps $dt= 9\tau/(250n)$.}\BibitemShut
  {Stop}%
\bibitem [{sup()}]{supboil}%
  \BibitemOpen
  \href@noop {} {}\bibinfo {note} {The drops in $\rho_{\rm atm}(\tilde{T})$
  shown in Figure \ref{fig:phase} occur in liquids that are superheated;
  $\tilde{T}_{\rm boil}$ measured in this way are above the equilibrium boiling
  temperatures.}\BibitemShut {Stop}%
\bibitem [{why()}]{whySast}%
  \BibitemOpen
  \href@noop {} {}\bibinfo {note} {Ref.\ \cite{sastry97} examined monodisperse
  Lennard-Jones systems. They did not report locally-crystalline IS for $\rho <
  \rho_S$, perhaps because the very small system sizes ($N=256$) employed in
  the majority of their simulations obscured the trends discussed
  here.}\BibitemShut {Stop}%
\end{thebibliography}

%

\end{document}